\documentclass[preprint2,usenatbib]{mnras}
\usepackage{graphicx}
\usepackage {amsmath,amssymb}
\usepackage{times}  
\usepackage{color}
\usepackage{xcolor}
\usepackage{aas_macros}
\usepackage{array}
\usepackage{ulem}

\newcommand{\Msun}{{\rm M}_{\odot}}

\title[Low surface brightness galaxies in IllustrisTNG]{The formation of low surface brightness galaxies in the IllustrisTNG simulation}

\author[P\'erez-Monta\~no, et al.]
{
	\parbox{18cm}{
		Luis Enrique P\'erez-Monta\~no,$^{1}$\thanks{E-mail: l.perez@irya.unam.mx}
        Vicente Rodriguez-Gomez,$^{1}$
        Bernardo Cervantes Sodi,$^{1}$
        Qirong Zhu,$^{2}$
        Annalisa Pillepich,$^{3}$
        Mark Vogelsberger$^{4}$
        and Lars Hernquist$^{5}$
	}
	\vspace{0.3cm} \\ 
	$^{1}$ Instituto de Radioastronomía y Astrofísica, Universidad Nacional Autónoma de México, Antigua Carretera a Pátzcuaro \# 8701, \\Ex-Hda. San José de la Huerta, Morelia, Michoacán, México C.P. 58089 \\
	$^{2}$ McWilliams Center for Cosmology, Department of Physics, Carnegie Mellon University, Pittsburgh, PA 15213, USA \\
	$^{3}$ Max-Planck-Institut f\"{u}r Astronomie, K\"{o}nigstuhl 17, D-69117 Heidelberg, Germany \\
	$^{4}$ Department of Physics, Kavli Institute for Astrophysics and Space Research, Massachusetts Institute of Technology, Cambridge, MA 02139, USA \\
	$^{5}$ Harvard-Smithsonian Center for Astrophysics, 60 Garden Street, Cambridge, MA 02138, USA \\
}

\date{Accepted XXX. Received YYY; in original form ZZZ}

\pubyear{2022}

\begin{document}
\label{firstpage}
\pagerange{\pageref{firstpage}--\pageref{lastpage}}
\maketitle

\begin{abstract}
We explore the nature of low surface brightness galaxies (LSBGs) in the hydrodynamic cosmological simulation TNG100 of the IllustrisTNG project, selecting a sample of LSBGs ($r$-band effective surface brightness $\mu_r > 22.0$ mag arcsec$^{-2}$) at $z=0$ over a wide range of stellar masses ($M_{\ast} = 10^{9}$--$10^{12}$ M$_\odot$). We find LSBGs of all stellar masses, although they are particularly prevalent at $M_{\ast} < 10^{10}$ M$_\odot$. We show that the specific star formation rates of LSBGs are not significantly different from those of high surface brightness galaxies (HSBGs) but, as a population, LSBGs are systematically less massive and more extended than HSBGs, and tend to display late-type morphologies according to a kinematic criterion.
At fixed stellar mass, we find that haloes hosting LSBGs are systematically more massive and have a higher baryonic fraction than those hosting HSBGs. We find that LSBGs have higher stellar specific angular momentum and halo spin parameter values compared to HSBGs, as suggested by previous works. We track the evolution of these quantities back in time, finding that the spin parameters of the haloes hosting LSBGs and HSBGs exhibit a clear bifurcation at $z \sim 2$, which causes a similar separation in the evolutionary tracks of other properties such as galactic angular momentum and effective radius, ultimately resulting in the values observed at $z =$ 0. The higher values of specific stellar angular momentum and halo spin in LSBGs seem to be responsible for their extended nature, preventing material from collapsing into the central regions of the galaxies, also causing LSBGs to host less massive black holes at their centres.

\end{abstract}




\section{Introduction}
The existence of low surface brightness galaxies (LSBGs) was first speculated by \citet{Disney76} after \citet{Freeman70} reported that the majority of the galaxies in his sample shared very nearly the same central surface brightness in the $B-$band, with values around $\mu_B = $ 21.65 mag arcsec$^{-2}$. \citet{Disney76} argued that this finding was due to a severe selection bias, and that there should be galaxies with lower surface brightness, comparable to that of the night sky. Later observational work \citep{McGaugh95a} confirmed the existence of such galaxies, popularizing the term LSBG for galaxies fainter than a specified central surface brightness value.

 LSBGs constitute an important fraction of extragalactic sources, representing $\sim$ 30--50\% of the overall galaxy population \citep{McGaugh95a, ONeil03}. With the development of new observational techniques, larger telescopes and more sensitive instruments, it is now possible to study a large variety of this kind of objects, from the giant LSBGs such as Malin 1 \citep{Bothun87} to the more abundant dwarf LSBGs \citep{Impey97}.

LSBGs are primarily bulgeless, late-type galaxies, typically with bulge-to-disc luminosity ratios below 0.1 \citep{McGaugh95b}. The fraction of barred LSBGs is low, and when they display a bar it is usually less prominent and shorter than in HSBGs \citep{Honey16, Cerv17}. Among LSBGs, only $\sim$5-20\% host an active galactic nucleus (AGN), while the AGN fraction of HSBGs reaches up to 50\% \citep{Impey96, Galaz11}. In addition, LSBGs are bluer than HSBGs, more gas-rich \citep{Huang14,Du15}, display lower metallicities, typically $Z < $ 0.003 \citep{deBlok98a,deBlok98b,deNaray04}, and have low dust masses \citep{Hinz07,Rahman07}. Finally, LSBGs are found to be dark matter (DM) dominated galaxies at all radii \citep{deBlok01,McGaugh01,PerezMontano19}, typically showing slowly rising rotation curves.

One classical framework invoked to understand the formation of LSBGs postulates that they are formed at the centres of high angular momentum dark matter haloes \citep{Dalcanton97, Jimenez98, Hdz98, Mo98, Boissier03}. The origin of the angular momentum is usually explained under the Tidal Torque Theory, first suggested by \cite{Hoyle49} and later developed by \citet{Peebles69}, where protogalaxies acquire their angular momentum by tidal torques exerted by neighbouring overdensities of other protogalaxies in the early Universe. Within this framework, LSBGs are expected to form in haloes of high angular momentum as specified by their spin parameter $\lambda$, which is believed to determine some properties like the scale-length of the discs and their stellar surface density \citep{Fall80,Hdz98,Mo98,HdzCerv06,Cervantes13}. Under this hypothesis, simplified models have been constructed, successfully reproducing the structural and chemical evolution of observed samples of LSBGs \citep{Jimenez98, Boissier03, PerezMontano19, Salinas21}. Similarly, using a suite of idealized hydrodynamic simulations of disc galaxies, \citet{KimLee13} concluded that the angular momentum of the host halo was the determinant parameter to form galaxies with low surface brightness.

In order to understand how LSBGs arise within a cosmological context, including the effects of mergers, environment, and more realistic and diverse formation histories, the use of cosmological hydrodynamical simulations has gained interest in recent years (see \citealt{Somerville15, Vogelsberger20}). For example, \citet{Zhu18} reported an analogue of Malin 1 \citep{Bothun87}, the archetypal example of a giant LSBG, using the IllustrisTNG simulation \citep{Marinacci18, Naiman18, Nelson18, Pillepich18, Springel18}, exploring its formation history and finding a good agreement with the observational parameters of this galaxy, such as its extended disc, metallicity, and rotation curve. \cite{Zhu18} also highlighted the importance of galaxy mergers in the formation of \textit{giant} LSBGs, finding that their Malin 1-like simulated galaxy underwent a recent interaction with a pair of `intruding' galaxies, which triggered gas cooling from the hot halo gas, forming an extremely extended stellar disc.

Similarly, with a sample of galaxies derived from EAGLE \citep{Schaye15,Crain15,McAlpine16}, \citet{Kulier20} found that LSBGs and HSBGs  are formed within DM halos with similar mass, but being the LSBGs the ones that have a higher baryonic budget compared to HSBGs, mostly composed of non-star-forming gas. They also found that mergers and the accretion of stellar mass have an important contribution to the formation of extended stellar discs around massive LSBGs. \citet{Martin19} found with the aid of the HorizonAGN simulation \citep{Dubois14} that LSBGs and HSBGs are formed from the same population of objects, but LSBGs evolve faster, especially at $z \sim$ 1, where LSBGs have lost large amounts of gas due to ram-pressure stripping and tidal forces, affecting their evolution especially at lower redshifts. Finally, \citet{DiCintio19}, using a sample of 12 galaxies from the NIHAO zoom-in simulations \citep{Nihao}, showed that LSBGs are formed inside DM halos with $\lambda > $ 0.04, and their density profiles are shallower, as a result of the ejection of baryons by stellar feedback. They also studied the role that mergers play in the formation of LSBGs, finding that LSBGs are formed by the combination of two effects: the direction in which these mergers occur, and the direction in which the gas falls into these galaxies during their formation.

In this work, we employ the TNG100 simulation from the IllustrisTNG project to construct a synthetic sample of LSBGs at $z=$ 0, in order to study their global physical parameters. We compare them to a sample of HSBGs drawn from the same simulation, with the aim of identifying the key parameters responsible for their low surface brightness nature. We also examine the redshift evolution and star formation histories of these two galaxy populations in order to identify systematic differences in their formation mechanisms. This paper expands upon previous studies by considering a larger sample of LSBGs over a wider range of stellar masses, and by presenting a detailed exploration of the evolutionary pathways followed by LSBGs and HSBGs. Additionally, we minimize possible selection effects due to the dependence of most galactic properties on stellar mass by always contrasting LSBGs and HSBGs within narrow stellar mass intervals.

This paper is organized as follows. In Section \ref{sec:Sample} we describe the simulation employed in this work. In Section \ref{sec:Results_local} we present the main results of our work, describing the global statistics of our sample, the distributions of different physical parameters, and a detailed comparison of these physical parameters between LSBGs and HSBGs at $z=0$. Section \ref{sec:lsb_evol} presents the redshift evolution and star formation histories of our galaxies. Finally, in Section \ref{sec:Conclusions} we provide a general discussion and summarize our results.

\begin{figure*}
\centering
\begin{tabular}{cc}
\includegraphics[width=0.46\textwidth]{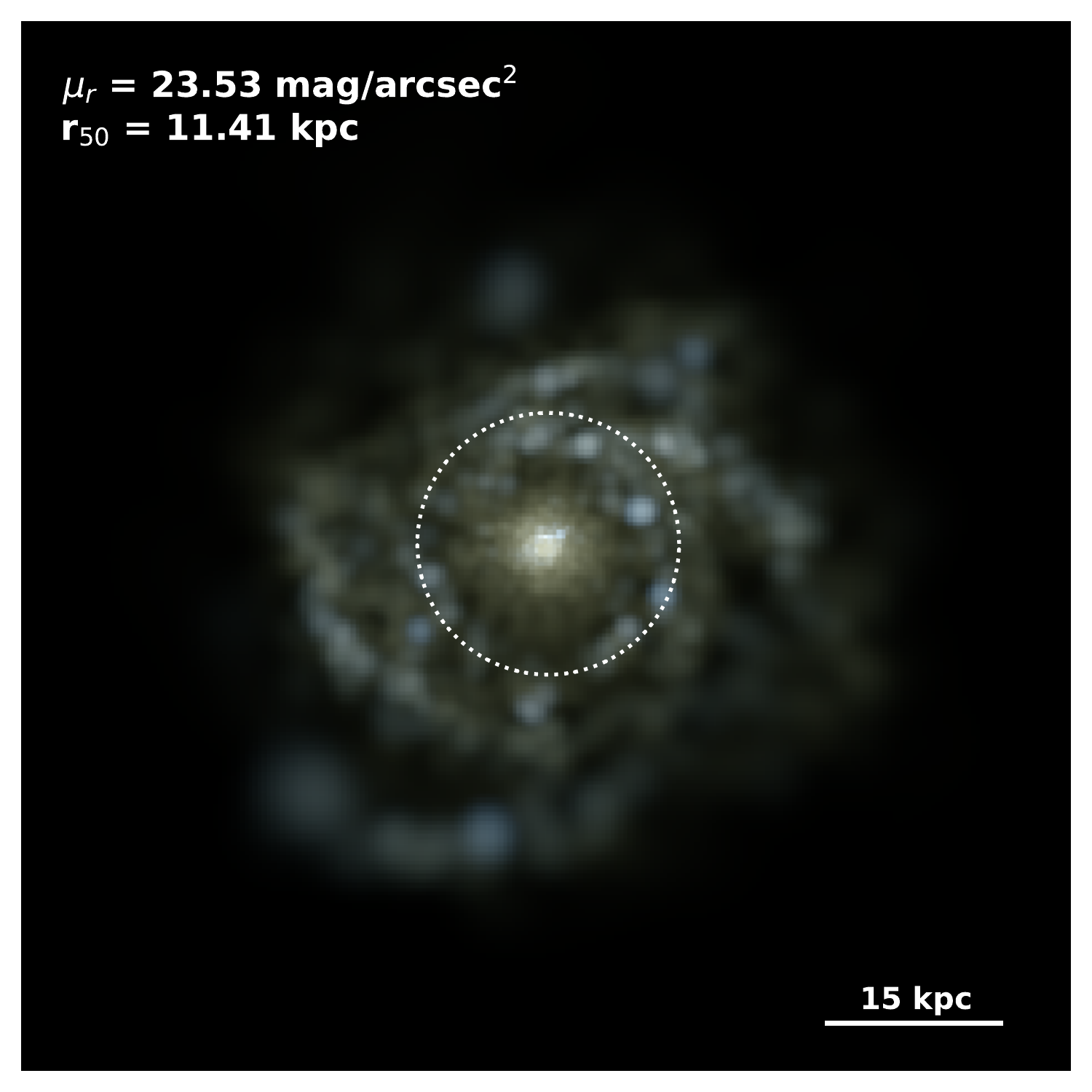} & 
\includegraphics[width=0.46\textwidth]{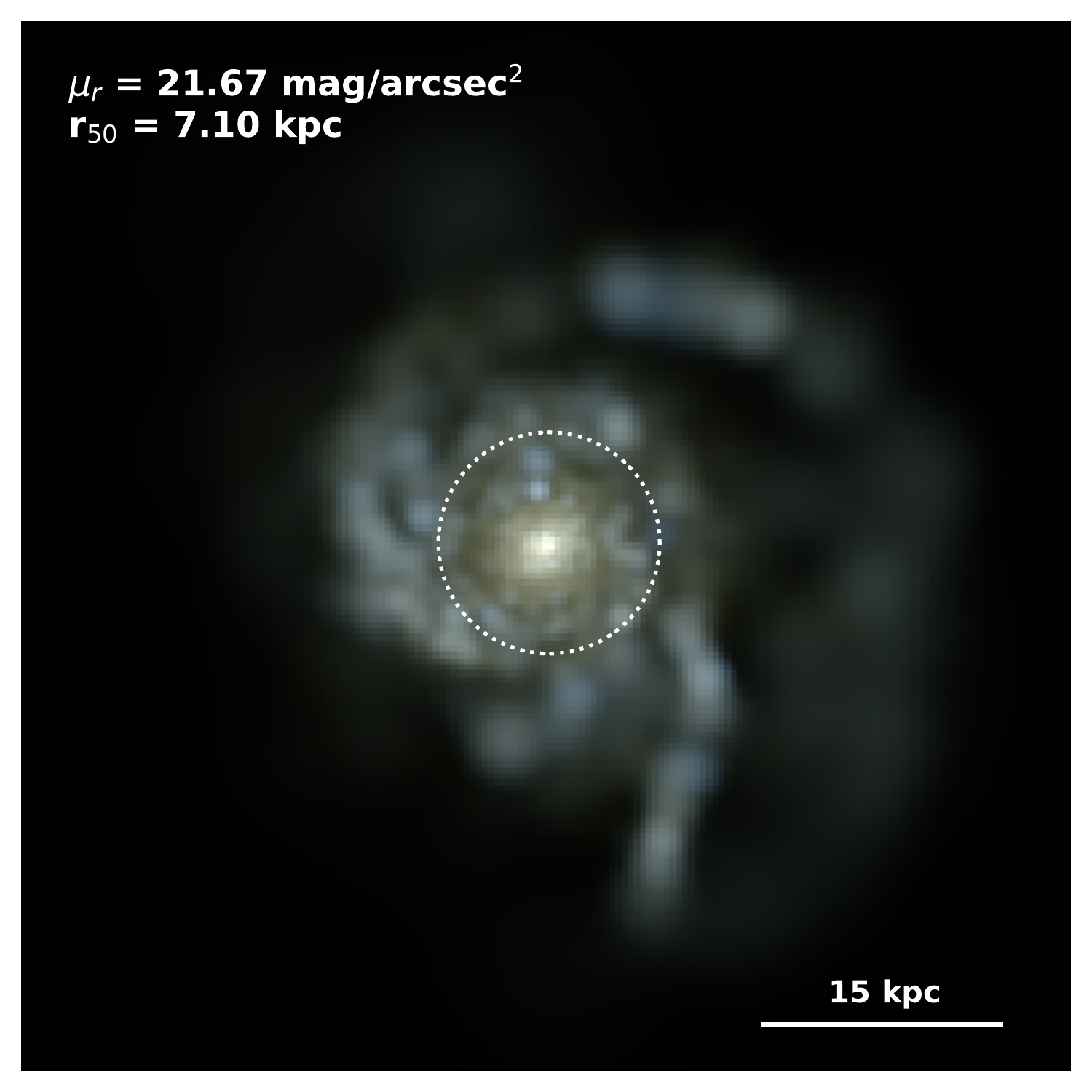} \\
\end{tabular}
\caption{Comparison between a randomly selected LSBG (left) and an HSBG (right) of approximately the same stellar mass, $\sim 10^{10}$ M$_{\odot}$. The upper-left corner of each panel shows the values of the effective surface brightness ($\mu_r$) and the $r$-band half-light radius ($r_{50}$). Dotted-line circle encloses the region delimited by $r_{50}$. We can observe that, despite having a similar mass, the LSBG spreads its stellar content over a larger area and therefore its surface brightness is lower.}
\label{fig:lsb_vs_hsb}
\end{figure*}

\section{Simulated Sample}
\label{sec:Sample}

\subsection{The TNG simulation}
\label{sec:sample}
The IllustrisTNG project (hereafter TNG, \citealt{Nelson18,Pillepich18,Springel18,Naiman18,Marinacci18,Nelson19a,Nelson19b,Pillepich19}) is a set of 18 magneto-hydrodynamical cosmological simulations that constitute an upgraded version of the original galaxy formation model used in the Illustris project \citep{Genel14,Vogelsberger14, Vogelsberger14b,Nelson15} that varies in size, resolution and complexity of the physical processes involved. The simulation follows the gas dynamics coupled with DM throughout a quasi-Lagrangian treatment, using the moving-mesh code \texttt{AREPO} \citep{Arepo10,Pakmor16}. \texttt{AREPO} allows a dynamical adaptive discretization, in which the cells of the mesh that allow the solution of the hydrodynamical equations move with the gas. The main advantages of TNG consist in model improvements over those of the original Illustris project \citep{Weinberger17,Pillepich18a}, as well as a wider range of simulation volumes and resolutions.

In this work we use the highest-resolution version of the TNG100 simulation, which follows the evolution of $1820^3$ DM particles and approximately $1820^3$ baryonic resolution elements (gas cells and stellar particles) over a periodic volume of $\sim$100 Mpc per side. The mass of each DM particle is $7.5 \times 10^6 \, \Msun$, while the average mass of the baryonic resolution elements is $1.4 \times 10^6 \, \Msun$. The gravitational softening length of the DM and stellar particles, which roughly determines the spatial resolution of the simulation, is 0.74 kpc at $z=$ 0. The output of the simulation consists of 100 snapshots running from $z=$ 20 to $z=$ 0, along with their associated halo catalogs and merger trees.

The simulation uses a Friends-of-Friends algorithm (FoF, \citealt{Davis85}) to identify the DM halos. Each halo contains subhalos, which are identified as overdense, gravitationally bound substructures with the \textsc{SUBFIND} algorithm \citep{Springel01, Dolag09}. Once the haloes and subhaloes have been identified for every snapshot, merger trees are constructed with the \textsc{sublink} algorithm \citep{RodGom15}.

Throughout this paper, we adopt the cosmological parameters derived from \citet{Planck16}, with $\Omega_m=$ 0.31, $\Omega_\Lambda=$ 0.69, $\Omega_b=$  0.0486 and $h=$ 0.677.

\subsection{Sample construction}
\label{sec:sample_const}

In order to have a sufficiently large number of objects for our study, which are at the same time sufficiently resolved, our sample includes all objects with $M_{*}$ $> 10^{9}$ M$_\odot$ at $z=0$. In order to segregate the sample into LSBGs and HSBGs, we compute the mean central surface brightness in the $r-$band of the Sloan Digital Sky Survey (SDSS) as in \citet{Zhong08} and \citet{Bakos12},  given by

\begin{equation}
	\label{eq:mu_x}
	\mu_r = m_r +2.5 \log{(\pi r_{50,r}^2)},
\end{equation}

\noindent with $m_r$ being the apparent magnitude within the projected half-light radius $r_{50,r}$. All galaxies are analyzed as viewed face-on, with the angular momentum vector of the stellar component orientated along the line of sight, and assuming that the galaxies are located at $z=0.0485$. No dust attenuation is considered, with fluxes calculated using the stellar population synthesis models by \citet{BC03}. \citet{Kulier20} demonstrated that dust attenuation does not play a significant role on the properties of LSBGs, especially when viewed face-on.

The full sample consists of 22,554 galaxies, including 5,814 LSBGs and 16,740 HSBGs, segregated by a limiting surface brightness of $\mu_r = 22.0$ mag arcsec$^{-2}$ (adopting the selection criterion from \citealt{DiCintio19}), which, in our simulated sample, roughly corresponds to the classical value of $\mu_B = 21.65$ mag arcsec$^{-2}$ \citep{Freeman70}. It is important to point out that, in the present work, our attempt is not a quantitative comparison with previous theoretical or observational works, but a qualitative one, as the segregation criteria to define HSBGs and LSBGs may vary greatly. For instance, in this work we choose the central surface brightness within the half-light radius in the $r-$band of the SDSS, including all morphological types; while other works may choose among the central surface brightness, the central surface brightness within a standard isophote or within the scalelength radius from an exponential light profile, all of them measured in different bands and with the inclusion, or not, of certain morphological types.

Figure \ref{fig:lsb_vs_hsb} shows synthetic images of two galaxies in our sample, obtained following the prescription by \citet{RodGom19}. The left-hand panel corresponds to a randomly selected LSBG, while the right-hand panel shows an HSBG with similar stellar mass. We can clearly note that galaxies classified as LSBs are more diffuse and extended objects than HSBGs.

In order to distinguish between spiral and elliptical galaxies, we consider a kinematic criterion and adopt the $\kappa_{\mathrm{rot}}$ parameter \citep{Sales10, RodGom17}, defined as the ratio between the stellar kinetic energy invested into circular motion along the azimuthal component, $K_{\mathrm{rot}}$, and the total stellar kinetic energy, $K$, i.e. 

\begin{equation}
\label{eq:kappa}
	\kappa_{\rm rot} = \frac{K_{\rm rot}}{K}.
\end{equation}

\begin{figure}
\centering
\includegraphics[width=0.5\textwidth]{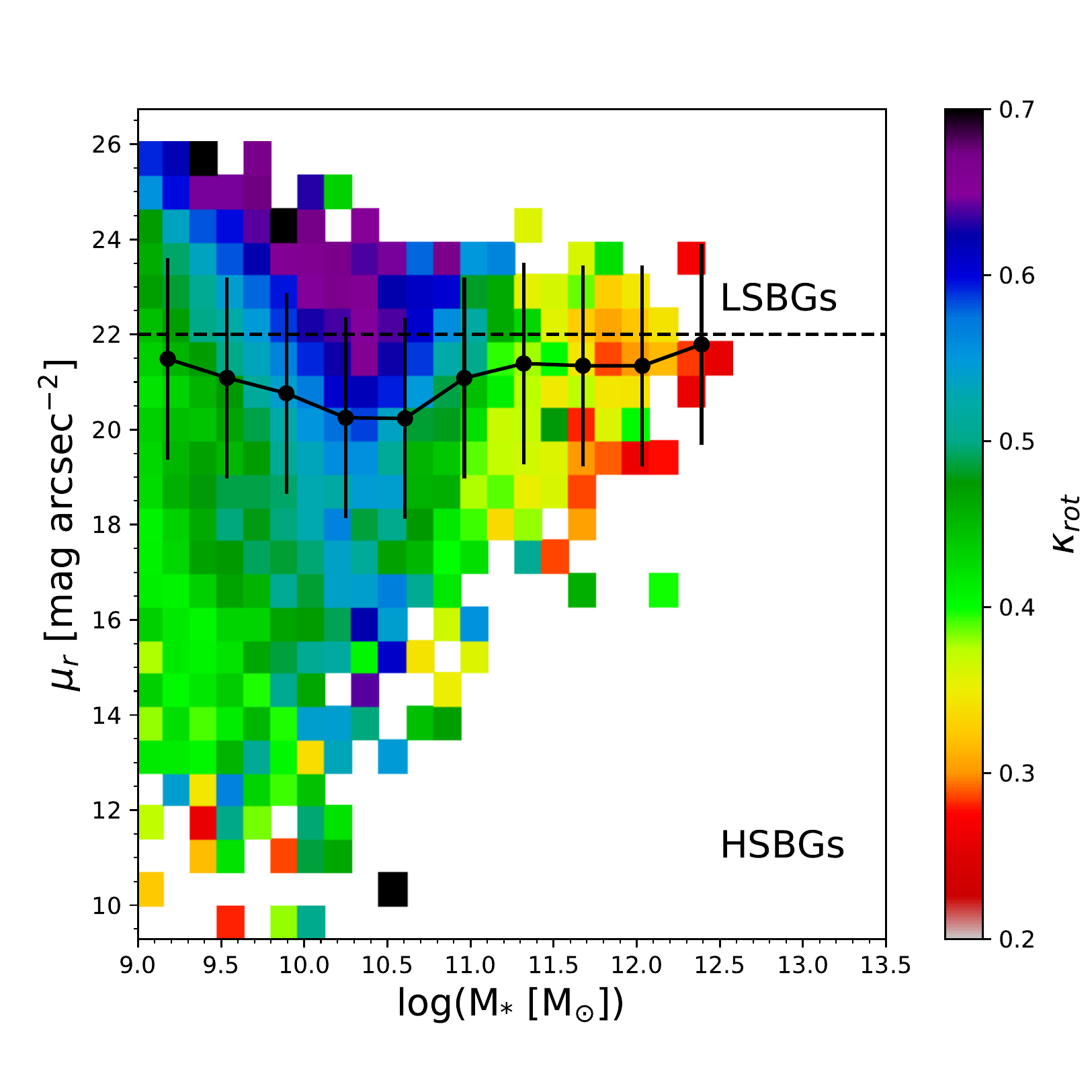} \\
\includegraphics[width=0.45\textwidth]{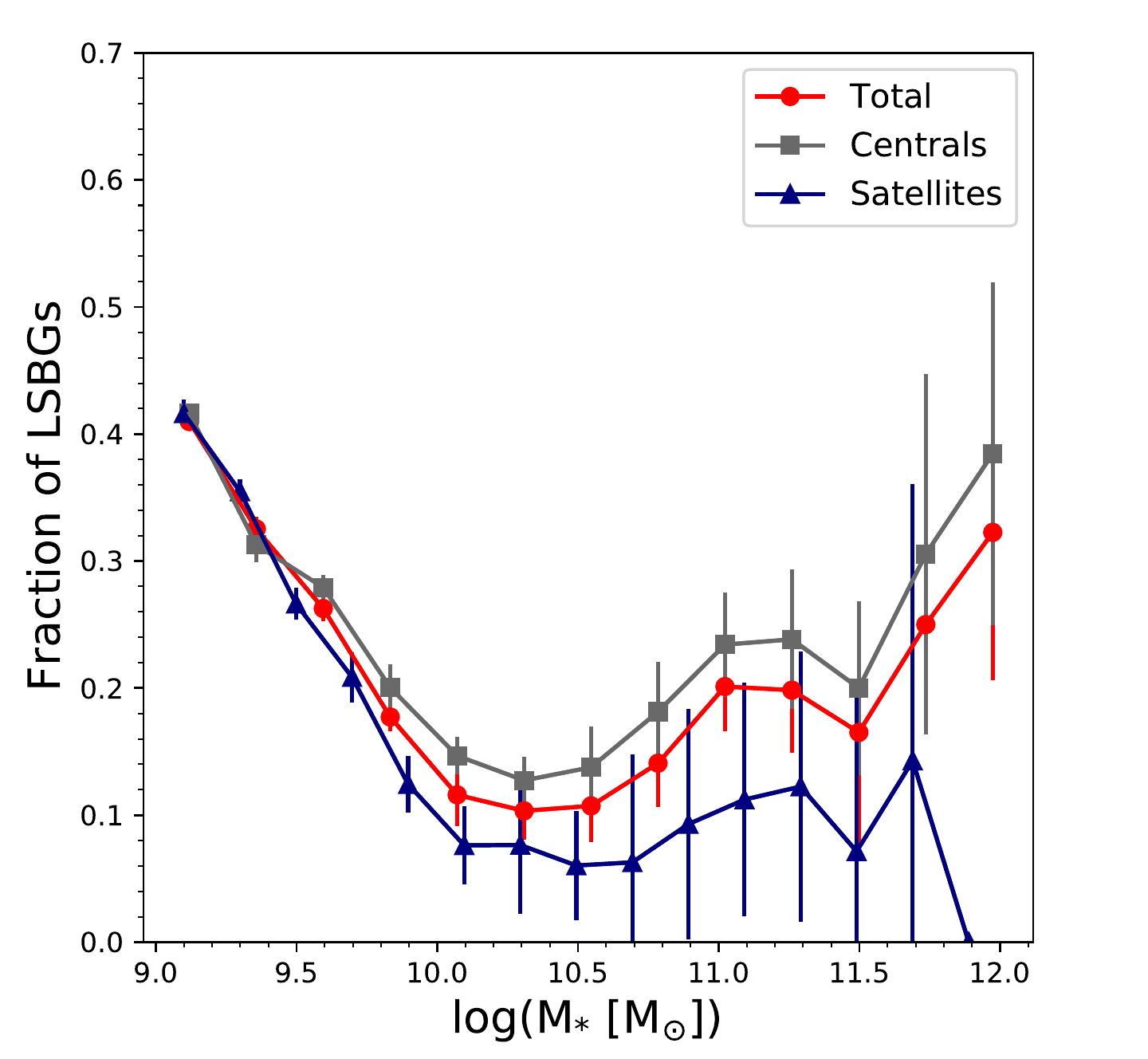} \\
\caption{\textit{Top Panel:} Surface brightness $\mu_r$ as a function of the stellar mass of the galaxies in our synthetic sample, colour-coded by their $\kappa_{\rm rot}$ value. We observe that the LSBG population consists mainly of spiral galaxies at stellar masses lower than 10$^{11}$ M$_\odot$. \textit{Bottom panel:} Fraction of LSBGs as a function of stellar mass, segregated into centrals (gray squares), satellites (navy triangles), and total (red dots).}
\label{fig:mu_vs_Mstar_kappa}
\end{figure}

In general, $\kappa_{\rm rot}$ values above 0.5 can be considered to correspond to late-type galaxies, although we emphasize that we do not make any morphological selection throughout the paper. The top panel of Fig. \ref{fig:mu_vs_Mstar_kappa} shows the effective surface brightness (measured within one half-light radius) computed in the $r$-band as a function of stellar mass, with the color code representing the median value of $\kappa_{\rm rot}$ of the galaxies within each $\mu_r$ -- log($M_{*}$) bin. The dashed line at $\mu_r=$ 22 mag arcsec$^{-2}$ segregates our galaxies into LSBGs and HSBGs. The black points correspond to the median value of $\mu_r$ and the errorbars indicate the 16th to 84th percentile range of the $\mu_r$ distribution for the full sample. The upper panel of Fig. \ref{fig:mu_vs_Mstar_kappa} shows that, for a wide range of masses (9 $\lesssim$ log($M_{*}$) $\lesssim$ 11), LSBGs have a predominantly spiral morphology, while HSBGs show a wide range of morphologies, with ellipticals dominating at higher masses.

\begin{figure*}
\centering
\includegraphics[width=0.9\textwidth]{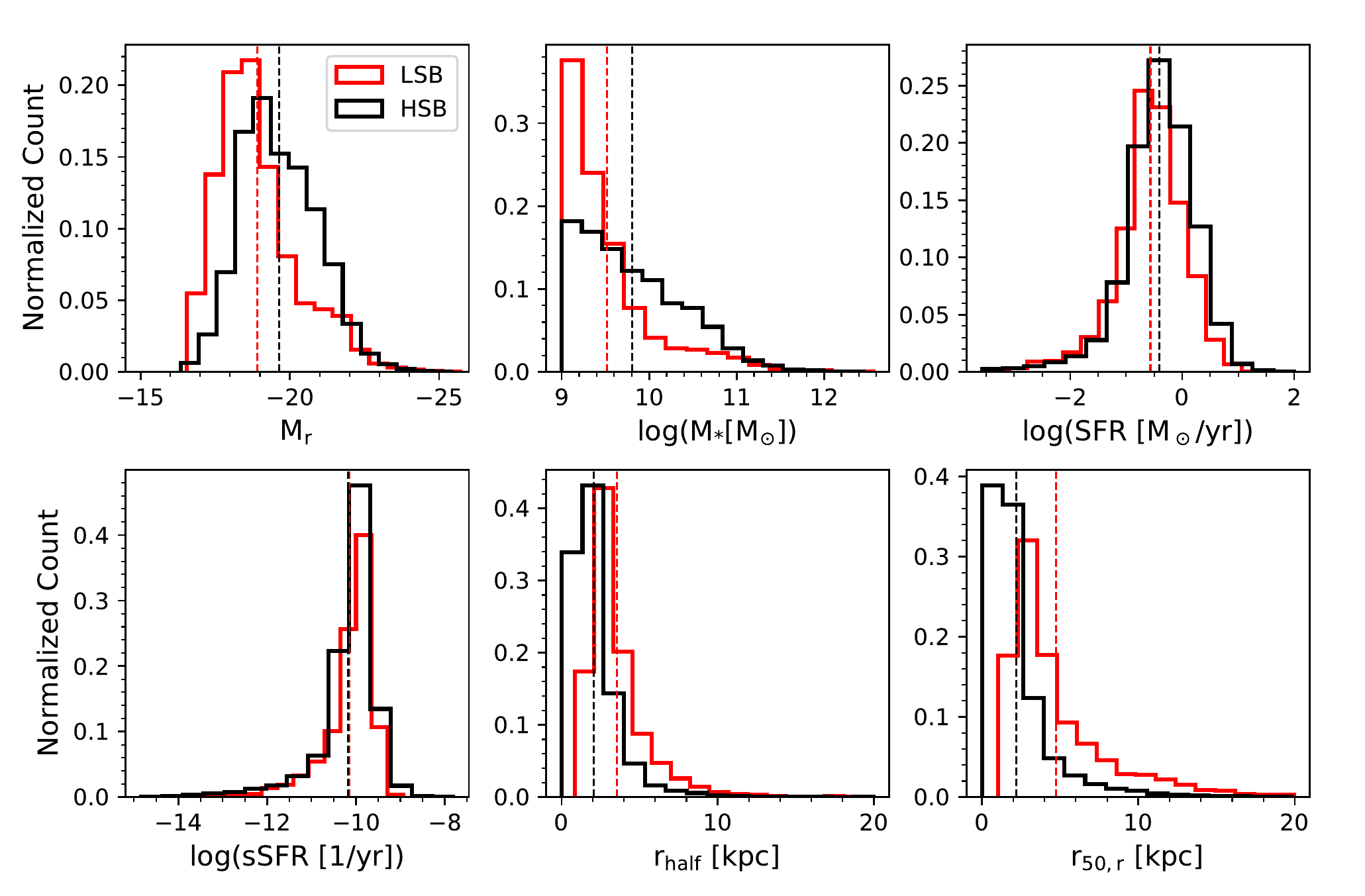} 
\caption{Distributions of the six main galaxy properties for LSBGs (red) and HSBGs (black) from the TNG100 simulation at $z = 0$, considering all galaxies with $M_{\ast} > 10^{9} \, \Msun$. The solid lines show the overall distribution of each galaxy property, while the vertical dotted lines show the corresponding mean values. This figure shows that LSBGs are systematically larger and less massive than HSBGs.}
\label{fig:galaxy_prop}
\end{figure*}

The bottom panel of Fig. \ref{fig:mu_vs_Mstar_kappa} shows the fraction of LSBGs at a given stellar mass, including a segregation between central and satellite galaxies. In all cases we note that there is an important fraction of LSBGs at low masses, which is expected given that these objects are mainly low-mass galaxies \citep{Dalcanton97,Galaz11}. However, an upturn in this trend is observed for central galaxies at $\sim 10^{10.5}$ M$_{\odot}$. A possible explanation for this behaviour is detailed in \citet{Kulier20}. The authors argued that stellar mass accreted via mergers contribute a significant fraction of the total stellar mass of these galaxies, but they extend over larger areas as a faint component, thus contributing to the LSBG population. 

It should be noted that LSBGs with stellar masses above 10$^{12}$ M$_{\odot}$ could have very extended components, particularly for the case of central galaxies of groups and clusters, where the inclusion of intra cluster light (ICL) could be determinant in their classification as LSBGs. In our sample we found 15 galaxies with $M_{\ast} > 10^{12} \, \Msun$ and $r_{50} >$ 30 kpc,\footnote{The 30 kpc threshold is a typical boundary employed to separate the ICL contribution from the main galaxy \citep{Schaye15,Pillepich18,Henden20}.} of which 11 of them are classified as LSBGs. Even if this population is very small compared with the full sample, this is consistent with the upturn found in the lower panel of Fig. \ref{fig:mu_vs_Mstar_kappa}. This could also partially explain the increasing LSBG fraction found by \cite{Kulier20} at high masses, considering that they computed the surface brightness within even larger apertures, with a possibly large ICL contribution at the massive end.
On the other hand, for galaxies with $M_{*}$ between 10$^{11.5}$ and 10$^{12}$ M$_{\odot}$, we find that the fraction of galaxies with $r_{50} >$ 30 kpc is less than 8\%, indicating that the rest of our galaxy classification is not particularly affected by the inclusion of the ICL. Therefore, the upcoming analysis will be limited to those galaxies with stellar masses lower than 10$^{12}$ M$_{\odot}$.

\begin{table}
	\centering 
	\begin{tabular}{c c c c c c} 
	\hline\hline 
	Type & $M_r$ &  log($M_*$)  & log(SFR) &  $r_{\rm half}$ & $r_{50,r}$ \\
	 -	 &	 $mag$ & M$_{\odot}$	& M$_{\odot}$ yr$^{-1}$ &  kpc & kpc \\ [0.5ex] 
	\hline 
	LSB & -18.906 & 9.515 & -0.567 & 3.562 & 4.72  \\ 
	HSB& -19.649 & 9.806 & -0.406 & 2.058 & 2.188 \\
	\hline
	$p$-value & 0.0 & 0.0 & $<$ 0.001 & 0.0 & 0.0 \\ 
	\hline
	\end{tabular}
	\caption{Mean values of the general galaxy properties for the synthetic LSBG and HSBG samples, along with the resulting $p$-values from performing a K-S test for each property.}
	\label{table:stel_prop} 
\end{table}

\section{Properties at low redshift}
\label{sec:Results_local}

\subsection{Global statistics}
\label{sec:global}
	We begin our analysis by exploring the distribution of six important physical quantities for our LSBGs and HSBGs sub-samples: the absolute $r$-band magnitude ($M_r$), the total stellar mass ($M_{\ast}$), the radius of the sphere containing half of the total stellar mass ($r_{\rm half}$), the projected effective radius containing 50\% of the luminosity in the $r$-band ($r_{50,r}$), the total star formation rate (SFR), and the specific star formation rate (sSFR). Figure \ref{fig:galaxy_prop} shows the distributions of these galaxy properties for LSBGs (red coloured) and HSBGs (black coloured), with their mean values reported in Table \ref{table:stel_prop}, including the corresponding $p$-value derived from a Kolmogorov–Smirnov (K-S) test to reject the null hypothesis that the distributions were extracted from the same underlying population.
	
	 Fig. \ref{fig:galaxy_prop} shows that LSBGs are on average less massive and fainter than HSBGs, and that HSBGs are less extended than LSBGs, quantified by their half-mass and half-light radii. On average, LSBGs have lower SFRs as a consequence of their lower masses. However, the sSFR distributions are almost indistinguishable from each other, which indicates that both galaxy types have similar levels of star formation activity.
	
\begin{figure*} 
\centering
\includegraphics[width=0.9\textwidth]{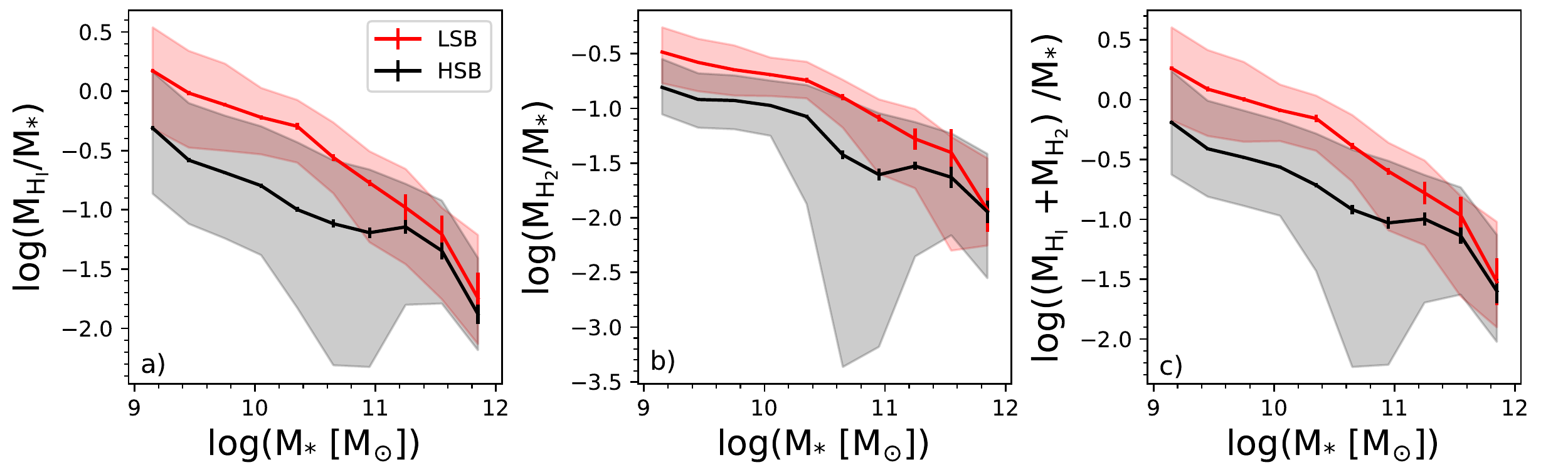} \\
\includegraphics[width=0.9\textwidth]{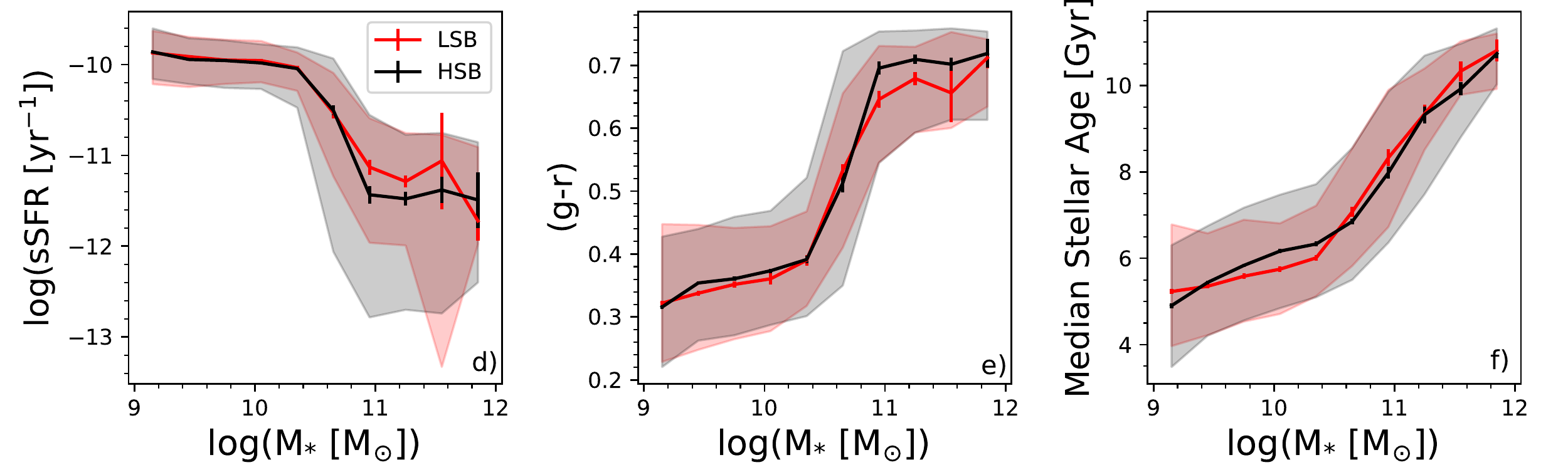}  \\
\caption{\textit{Upper panels}: Atomic (a), molecular (b) and neutral (c) hydrogen gas content as a function of $M_{*}$ for central LSBGs (red) and HSBGs (black) at $z = 0$. \textit{Lower panels}: Median values of the sSFR (d), color (e) and median age of the stars (f) as a function of $M_{*}$. Solid lines correspond to the median value, while the shaded regions enclose the interval between the 16th and 84th percentiles. Errorbars represent the dispersion around the median obtained from a bootstrap re-sampling algorithm. The following plots throughout this work follow the same format. We can see from this plot that LSBGs present systematically higher gas fractions than HSBGs, with no significant differences between their sSFR, color, and stellar population ages. We note, however, that the sharp transitions at $\log(M_{\ast} / \Msun) \sim 10.5$ in the distributions of color and sSFR are simply a reflection of the bimodal nature (at fixed stellar mass) of those distributions.}
\label{fig:gasSFR}
\end{figure*}

\subsection{Gas content and star formation}
\label{sec:gasSFR}
	 	 From this section forward, we will focus exclusively on central galaxies, given that for these galaxies the mass and spin of the parent DM halo are well-defined properties that have a direct effect on their formation. In our study, the central galaxy of a group is determined by the \textsc{subfind} algorithm, and usually corresponds to the most massive object. With this convention, our sub-samples consist of 7,240 LSBGs and 11,779 HSBGs with stellar masses above $10^{9}$ M$_{\odot}$.

        Given that the stellar mass distributions of LSBGs and HSBGs are different (Fig. \ref{fig:galaxy_prop}), we explore the differences between the two populations at fixed stellar mass, in order to control for any correlation of the particular property we are analyzing with $M_{\ast}$.
     	 	
	 	In the upper panels of Fig. \ref{fig:gasSFR} we show the atomic, molecular and neutral gas content (left-hand, middle and right-hand panel, respectively), obtained directly from  the TNG post-processed data \citep{Diemer18,Diemer19} following a \citet{GK11} model, for all the galaxies in the sample as a function of their stellar mass. Solid lines correspond to the median value of the properties presented, whereas the shaded regions enclose the interval between the 16th and 84th percentiles. Furthermore, the errorbars represent the dispersion of the median obtained from a bootstrap re-sampling algorithm, in which we performed a thousand random realizations derived from our original data set. We will keep this convention for subsequent figures. For galaxies with stellar masses lower than $10^{11.5}$ M$_\odot$, we clearly note that at a fixed stellar mass, LSBGs present higher gas content than their HSB counterparts, in agreement with previous observational studies (e.g. \citealt{ONeil00,Huang12}) that report a higher fraction of mass in gas of their total baryonic budget. More recently, \citet{Lutz18} demonstrated the existence of a link between angular momentum and HI gas content, finding that HI gas-rich galaxies tend to live in halos with high angular momentum, as is the case of the LSBG population in our sample. We will investigate the link between galactic angular momentum and LSBGs in more detail in Section \ref{sec:angmom}.

To complement these results, on the lower panels of Fig. \ref{fig:gasSFR} we present three properties related to the star formation history: the specific star formation rate sSFR $=$ SFR/$M_{*}$, the integrated colour $(g-r)$ and the median stellar age of the star populations weighted according to the stellar mass \citep{RodGom16}, for which the sample does not exhibit a clear distinction between LSBGs and HSBGs. 
 
 \begin{figure} 
\centering
\includegraphics[width=0.425\textwidth]{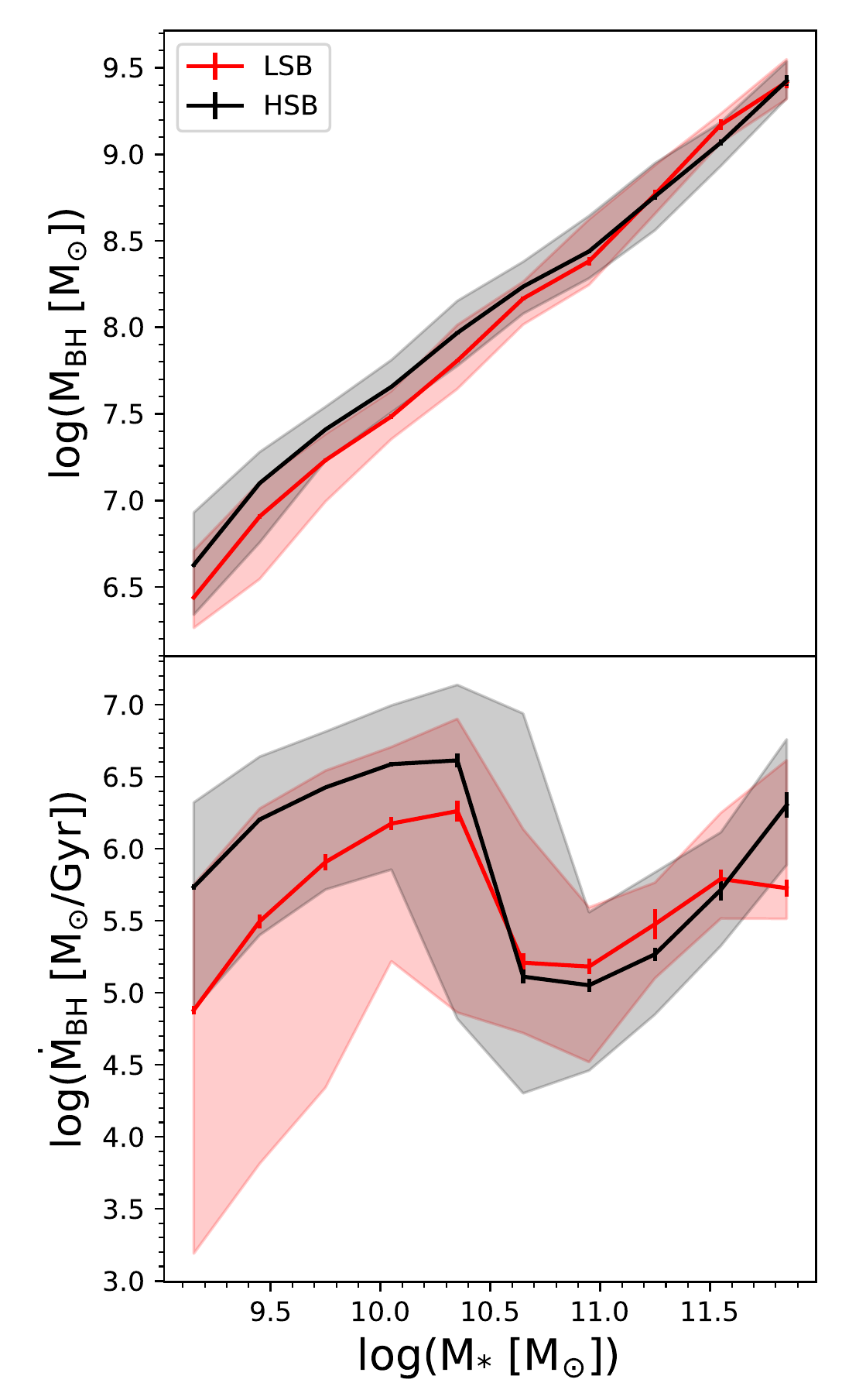} 
\caption{Median values of the BH mass (top) and BH accretion rate (bottom) of the galaxies in our sample. The colour convention is the same as in Fig. \ref{fig:gasSFR}. This figure shows that LSBGs host less massive BHs and have lower BH accretion rates than HSBGs (at fixed stellar mass), according to the IllustrisTNG model. Similarly to the color and sSFR distributions (Fig. \ref{fig:gasSFR}), the sharp transition at $\log(M_{\ast} / \Msun) \sim 10.5$ is due to a bimodality in the distribution of $\dot{M}_{\rm BH}$ (at fixed $M_{\ast}$) rather than a continuous trend.}
\label{fig:bhs}
\end{figure}

 \subsection{Black holes and nuclear activity}
 \label{sec:agns} 
 
    Having compared the gas content and star formation activity of LSBGs and HSBGs in our simulated samples, we proceed to explore the effects of AGN feedback -- which can heat gas and suppress star formation -- in these types of galaxies, some of which are reported to present AGN activity \citep{Das06}. As a proxy for the amount of energy ejected by the AGN, we study the masses and accretion rates of the supermassive black holes (BHs) at the centres of LSBGs and HSBGs.
    
    The top panel of Fig. \ref{fig:bhs} shows the median value of $M_{\rm BH}$ as a function of the total stellar mass, showing that for galaxies with stellar mass $M_{*}\le 10^{10.5}$ M$_\odot$, the BH masses of LSBGs are systematically lower than those of their high-surface-brightness counterparts at a fixed stellar mass. This result is qualitatively consistent with observational studies by \citet{Subramanian16} and \citet{Saburova21}, who found that the BH mass estimation for LSBGs is substantially below the expectation from the velocity dispersion of the bulge component. We note, however, that the observational samples of \citet{Subramanian16} and \citet{Saburova21} are relatively small and not representative of the whole galaxy population. On the theoretical side, our results agree with those of \citet{RodGom22}, who found an anticorrelation between $M_{\rm BH}$ and galactic angular momentum (which is higher in LSBGs), also using the TNG100 simulation. 

 	On the bottom panel of Fig. \ref{fig:bhs}, we plot the median values of the BH accretion rate, $\dot{M}_{\rm BH}$, as function of stellar mass. It is clear from this that the accretion rate for low-mass LSBGs is lower than for HSBGs. Given that LSBGs present a higher fraction of gas available for accretion to the central black hole, this last result would seem counter-intuitive, but the higher angular momentum content exhibited by LSBGs (see Section \ref{sec:angmom}) would prevent the gas from flowing to the central regions of the galaxies and feed the BH. As we will show in Section \ref{sec:angmom}, LSBGs present a higher amount of angular momentum and a larger value of the halo spin parameter than HSBGs, which reduces the flow of gas and stars toward the inner regions of these galaxies, resulting in lower gas densities in the vicinity of the BH and therefore reducing the accretion rate given the Bondi accretion model \citep{Bondi52} implemented in TNG. At the same time, the higher BH masses lead to more efficient AGN feedback, disrupting late-time gas accretion. Since the gas accreted at late times typically has higher specific angular momentum (e.g. \citealt{Ubler14}), this implies that higher BH masses result in lower galactic angular momentum \citep{Genel15, RodGom22} and therefore favour the formation of HSBGs.
 	
 	The non-monotonic shape of the BH accretion rate found in Fig. \ref{fig:bhs} is explained by the different modes of AGN feedback implemented in the simulation \citep{Weinberger17, Zinger20}. For low-mass galaxies, where the difference between LSBGs and HSBGs is more noticeable, the `thermal' mode of AGN feedback is expected to be dominant, which is characterized by higher accretion rates modulated by the central gas density, which is lower for the LSBG population. On the other hand, the dominant mode of AGN feedback in high-mass galaxies is the so-called `kinetic' mode, which is characterized by lower accretion rates and which typically reaches a self-regulated state where the exact value of the local gas density does not play a primary role.
 	 
 	The fact that LSBGs present a lower occurrence of stellar bars \citep{Honey16,Cerv17} may also contribute to the low accretion rate and undersized BHs in these galaxies, as bars promote mass and angular momentum redistribution between the various components of the galaxies, transferring gas to the center that could be used as material to feed the central black hole. In a recent work, also using the TNG100 simulation, \citet{Rosas20} reported that the median mass of the supermassive BHs hosted by strongly barred galaxies is systematically higher than the ones hosted by unbarred galaxies, an indication that these kind of mechanisms prompted by stellar bars might be playing a role in the BH accretion process in our sample.

\begin{figure*} 
\centering
\includegraphics[width=0.9\textwidth]{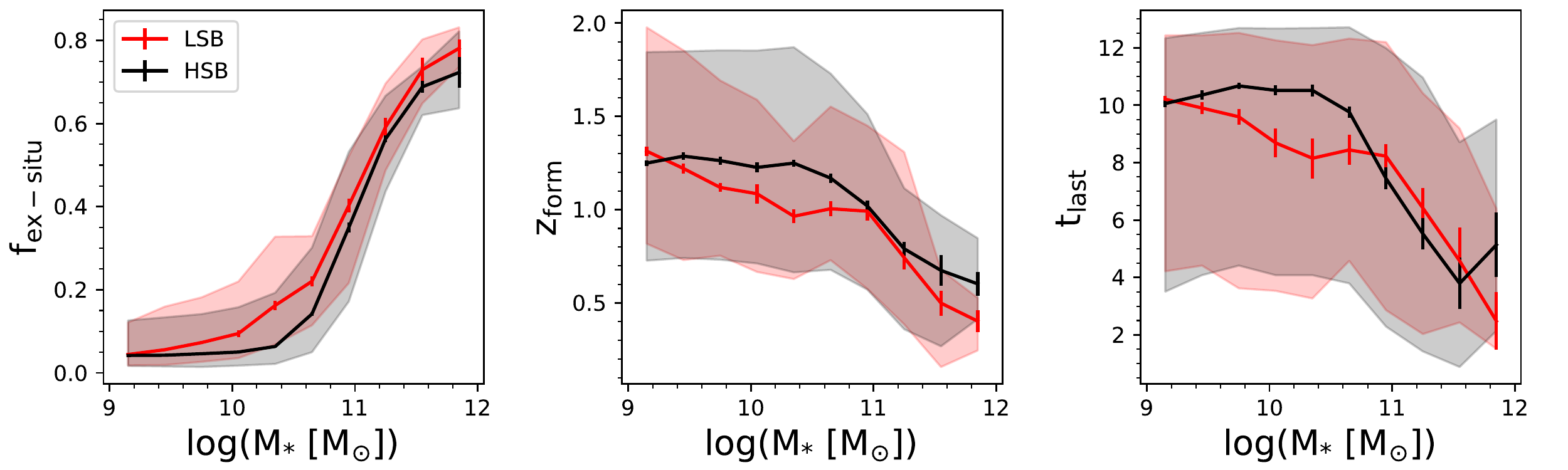} 
\caption{Median values of the fraction of the stellar mass accreted via mergers (left), redshift at which half of the final halo mass has assembled $z_{\rm form}$ (middle) and elapsed time since the last major merger $t_{\rm last}$ (right) for all the galaxies in our synthetic sample. This figure supports the idea that galaxy mergers favour the formation of LSBGs, but are probably not the dominant factor.}
\label{fig:history_prop}
\end{figure*}
 \subsection{Assembly and mergers}
 \label{sec:assamble}
 
 Fig. \ref{fig:history_prop} shows some properties related to the merging history of the galaxies in our synthetic sample, such as the \textit{ex situ} stellar mass fraction, defined as the fraction of stellar mass accreted via mergers (left-hand panel), the halo formation redshift $z_{\rm form}$, defined as the value of $z$ at which the DM halo gathered half of its final mass (middle panel), and the time since the last major merger (left-handed panel). All of these quantities were calculated as described in \cite{RodGom16}. We note that, in general, LSBGs present slightly higher stellar mass fractions accreted via mergers than HSBGs. We also observe that, for a restricted range of stellar masses, between $10^{9}$ and $10^{11}$ M$_\odot$, LSBGs experienced their last major merger more recently.
 
 These results are consistent with previous results obtained from simulations. \citet{Zhu18} found that the formation of a massive LSBGs such as Malin-1 is due to the encounter of a massive galaxy with a less massive gas-rich galaxy. \citet{DiCintio19} argue that certain merger configurations, such as coplanar mergers, are able to add angular momentum to the galaxy, leading to a decreasse in surface brightness. These works, in general, show that mergers could play an important role in the formation of LSBGs, depending on factors such as the direction of the merger, the angular momenta and alignment of the galaxies involved, their gas content, and their morphologies. In contrast, \citet{Martin19} found that both LSBGs and HSBGs undergo very few mergers at low redshift, and therefore mergers are unlikely to be the principal driver of LSBGs evolution over cosmic time. According to their findings, tidal perturbations and ram-pressure stripping may be the main mechanisms of LSBGs formation at low redshifts. As pointed out by \citet{RodGom16}, the ex-situ stellar component tends to have a more extended spatial distribution than that of in-situ stars, therefore contributing to the low surface brightness nature of ex-situ dominated galaxies, especially for the most massive ones, where the fractions of ex-situ stars reach $\sim$80\%. We will return to this point in Section \ref{sec:lsb_evol}.

\begin{figure*} 
\centering
\includegraphics[width=0.8\textwidth]{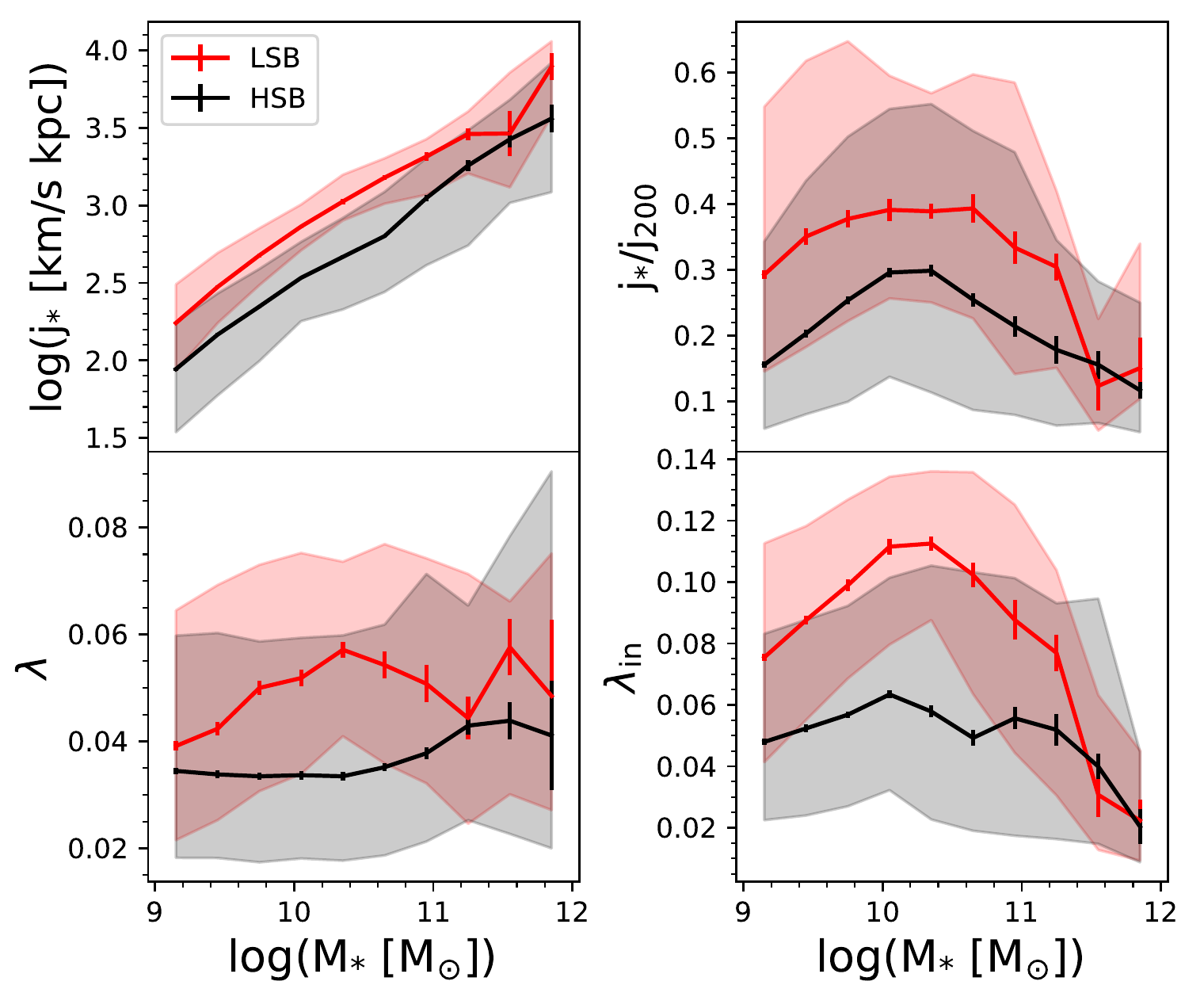}
\caption{Median values of the stellar specific angular momentum $j_*$ (top-left), angular momentum retention fraction $j_{*}/j_{200}$ (top-right), halo spin parameter $\lambda$ (bottom-left), and \textit{inner} halo spin parameter (bottom-right) as a function of stellar mass $M_{*}$, shown for all the galaxies in our sample. This figure shows that LSBGs have systematically higher stellar specific angular momentum than HSBGs (at fixed stellar mass), as a result of forming inside faster-spinning DM haloes and `retaining' a larger fraction of their specific angular momentum.}
\label{fig:angmom}
\end{figure*}

\subsection{Angular Momentum}
\label{sec:angmom}
	One of the major physical parameters determining the low surface brightness nature of these systems is the galactic angular momentum, which is related to the spin parameter of the parent DM halo (e.g. \citealt{Teklu15}; \citealt{RodGom22}). The top-left panel of Fig. \ref{fig:angmom} shows the median value of the stellar specific angular momentum $j_{*}$ as a function of the stellar mass $M_\ast$, with LSBGs presenting systematically higher values of $j_{*}$ across the entire stellar mass range of the sample, in agreement with findings from observational samples of nearby galaxies (\citealt{PerezMontano19}; \citealt{Salinas21}). Interestingly, \citet{Martin19} found that LSBGs and HSBGs have similar distributions of $j_{*}$. We attribute the difference between our result and theirs to the fact that we are comparing the HSBG and LSBG populations at fixed stellar mass, while they compared the whole galaxy populations without controlling for $M_\ast$.

	In the bottom-left panel of Fig. \ref{fig:angmom} we plot the median value of the halo spin parameter $\lambda$, computed using a proxy for the \citet{Peebles71} definition, proposed by \citet{Bullock01} as 

\begin{equation}
	\label{eq:spin_bullock}
	\lambda = \frac{j_{200}}{\sqrt{2} R_{200} v_{200}},
\end{equation}

\noindent where $j_{200}$ is the specific angular momentum of all the components within $R_{200}$, $R_{200}$ is the radius where the mean density is 200 times the critical density of the Universe, and $v_{200}$ is the circular velocity at $R=R_{200}$. From this panel, we note that LSBGs have higher values of $\lambda$, in agreement with the classical picture where LSBGs are formed within haloes with high values of the spin parameter (\citealt{Dalcanton97,Boissier03,KimLee13,PerezMontano19}). In order to exclude a possible `back reaction' effect of baryons on the DM halo spin, we have repeated this analysis by matching haloes from TNG100 to their counterparts from an analogous DM-only simulation, following the methodology of \citet{RodGom17}, finding no significant differences on the values of the total spin parameter shown in the bottom-left panel of Fig. \ref{fig:angmom}.

The bottom-right panel of Fig. \ref{fig:angmom} shows the median value for the spin parameter in the inner part of the halo $\lambda_{\rm in}$, computed within a radius equal to  $R=0.1 \; R_{200}$. This quantity is important because the stellar morphology of a galaxy is more closely connected to $\lambda_{\rm in}$ than to the value obtained for the whole configuration \citep{Zavala16}. The difference in $\lambda_{\rm in}$ between LSBGs and HSBGs appears to be more significant than the difference in $\lambda$, being LSBGs the ones with higher spin parameter. We note, however, that the $\lambda_{\rm in}$ values measured directly from a hydrodynamic simulation should be interpreted with caution, since the galaxy itself contributes significantly to this measurement. If we instead use $\lambda_{\rm in}$ values from matched haloes in TNG100-Dark, we find that the $\lambda_{\rm in}$ values decrease by a factor of $\sim$2--3 and the differences between LSBGs and HSBGs become comparable to those for the total halo spin, as shown on bottom-left panel of Fig. \ref{fig:angmom}. The behaviour of our sample is consistent with the results by \citet{Kulier20} using the EAGLE simulation.

	In the work of \citet{Fall83} and its subsequent revisions \citep{RomFall12, FallRom13}, it is shown that both elliptical and spiral galaxies follow roughly parallel tracks on the $j_{*} -  M_\ast$ plane, with the latter having a specific stellar angular momentum about 5 times larger than the former at a fixed stellar mass. Additionally, it has been found, using hydrodynamical simulations \citep{Kimm11,Stewart13,ZjupaSpringel17} that the specific angular momentum of the baryonic component of the galaxies is higher than that of the DM component. \citet{Teklu15} and \citet{Zavala16} found a correlation between $j_{*}$ and $j_{200}$, with a strong dependence with the morphology of the galaxies, with disc-dominated systems retaining a higher amount of angular momentum, as described by \citet{FallRom13}. Using the Illustris simulation, \citet{Genel15} found that, for spiral galaxies, the retention of angular momentum ($j_{\ast} / j_{200}$) is consistent with 100\%, whereas for early-type galaxies it is only 10-30\%. Similarly, using the TNG100 simulation, \citealt{RodGom22}) found retention fractions of 50--60\% for late types and 10--20\% for early types.
	
	With these results in mind, we explored whether there is a difference in the retention fraction of angular momentum between the two sub-samples in our study. The top-right panel of Fig. \ref{fig:angmom} shows the median value of $j_{*}/j_{200}$ as a function of the stellar mass. We note that throughout the full range of stellar masses explored, LSBGs retain a higher amount of angular momentum compared to HSBGs. Recently, \citet{RodGom22} showed that the angular momentum retention fraction is a strong function of $\kappa_{rot}$, while being mostly independent of stellar mass for the individual morphological types. Given that the LSBGs of our sample are mostly late-type (as seen in Fig. \ref{fig:mu_vs_Mstar_kappa}), the systematically higher values of $j_{*}/j_{200}$ are mostly a result of their morphology, rather than their intrinsic low surface brightness. The difference between LSBGs and HSBGs in the retained fraction of angular momentum by the stellar component, combined with the higher values of the halo spin parameter, could be determinant in the formation of LSBGs.

\begin{figure} 
\centering
\includegraphics[width=0.425\textwidth]{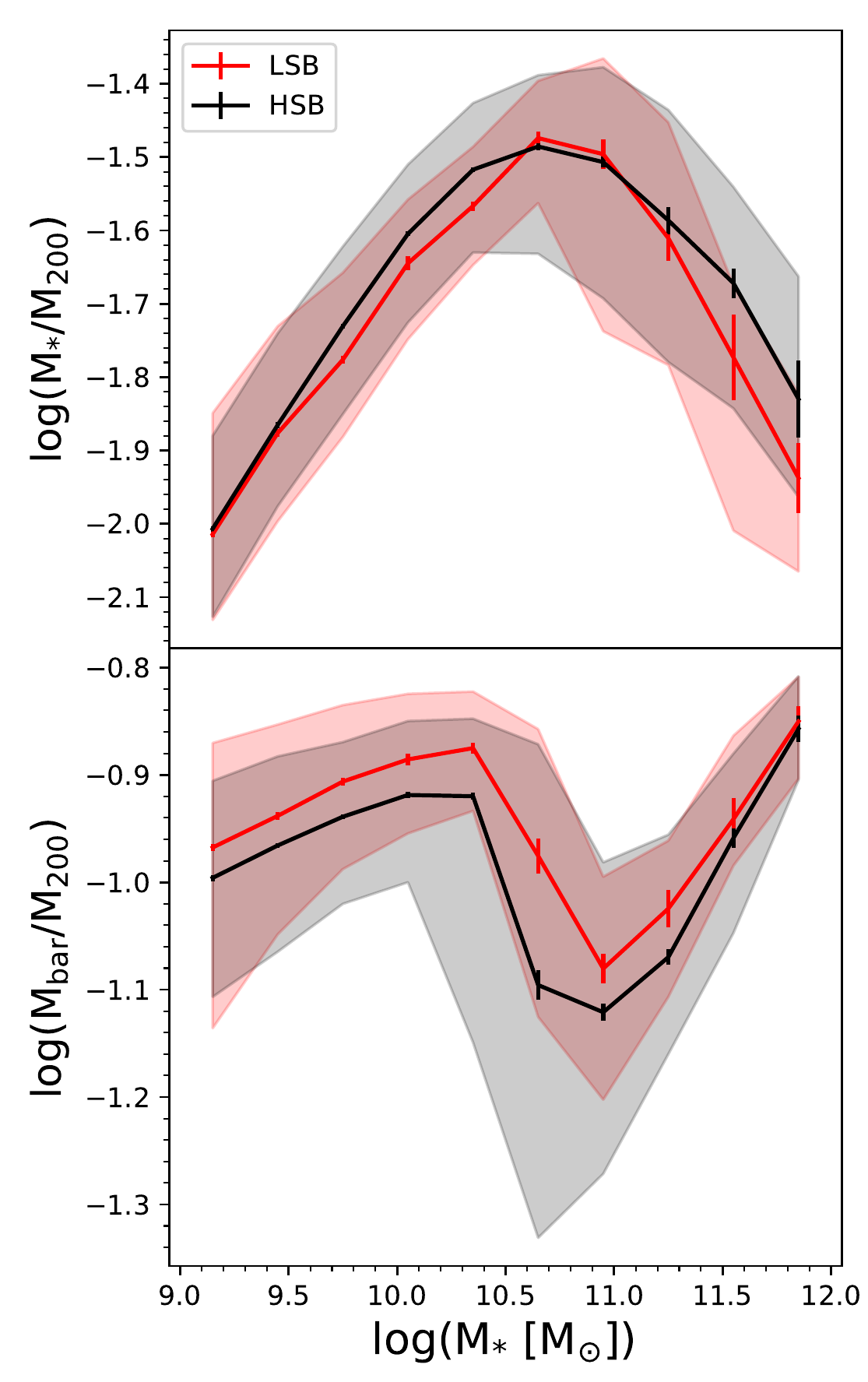} 
\caption{Median values of the stellar-to-halo (\textit{upper}) and baryonic-to-halo (lower) mass ratios as a function of stellar mass, shown for all central LSBGs and HSBGs at $z = 0$. At fixed stellar mass, LSBGs are formed within slightly more massive DM haloes with higher gas content than those hosting HSBGs. We also note that the joint distribution of $M_{\rm bar}/M_{200}$ and $M_{\ast}$ is bimodal, resulting in a sharp transition at $\log(M_{\ast} / \Msun) \sim 10.5$.}
\label{fig:halo_prop}
\end{figure}

\subsection{Halo Properties/Baryon Content}

\label{sec:halo} 
	We continue our study by analyzing the DM content of LSBGs, and how their baryonic and stellar content differs from that of HSBGs. We perform a similar exercise to that in the previous sections, studying the median values of the stellar-to-halo and baryonic-to-halo mass ratios, in a similar way as \citet{Kulier20}. We perform a similar analysis with our simulated galaxies by analyzing the stellar-to halo and baryonic-to-halo mass fractions. The upper panel of Fig. \ref{fig:halo_prop} shows the median values of $M_{*}/M_{200}$ (where $M_{200}$ is the total mass within $R_{200}$) as a function of stellar mass, in which we observe that LSBGs present a marginally smaller stellar-to-halo mass fraction compared to HSBGs.

	Analogously, the lower panel of Fig. \ref{fig:halo_prop} shows the median value of the baryonic-to-halo mass ratio, where $M_{\rm bar}= M_{*}+M_{\rm gas}$. We can observe that, for galaxies with stellar masses up to $10^{11}$ M$_\odot$, LSBGs have higher baryonic-to-halo mass ratio than HSB ones, as pointed out by \citet{Kulier20}. These authors found no difference in the stellar-to-halo mass ratio, but found a clear difference in the baryonic-to-halo mass ratio between LSBGs and HSBGs. They argued that, given that LSBGs are formed within haloes with similar mass at fixed stellar mass, the excess in the baryonic budget must consist of gas.

\begin{figure*} 
\centering
\includegraphics[width=0.9\textwidth]{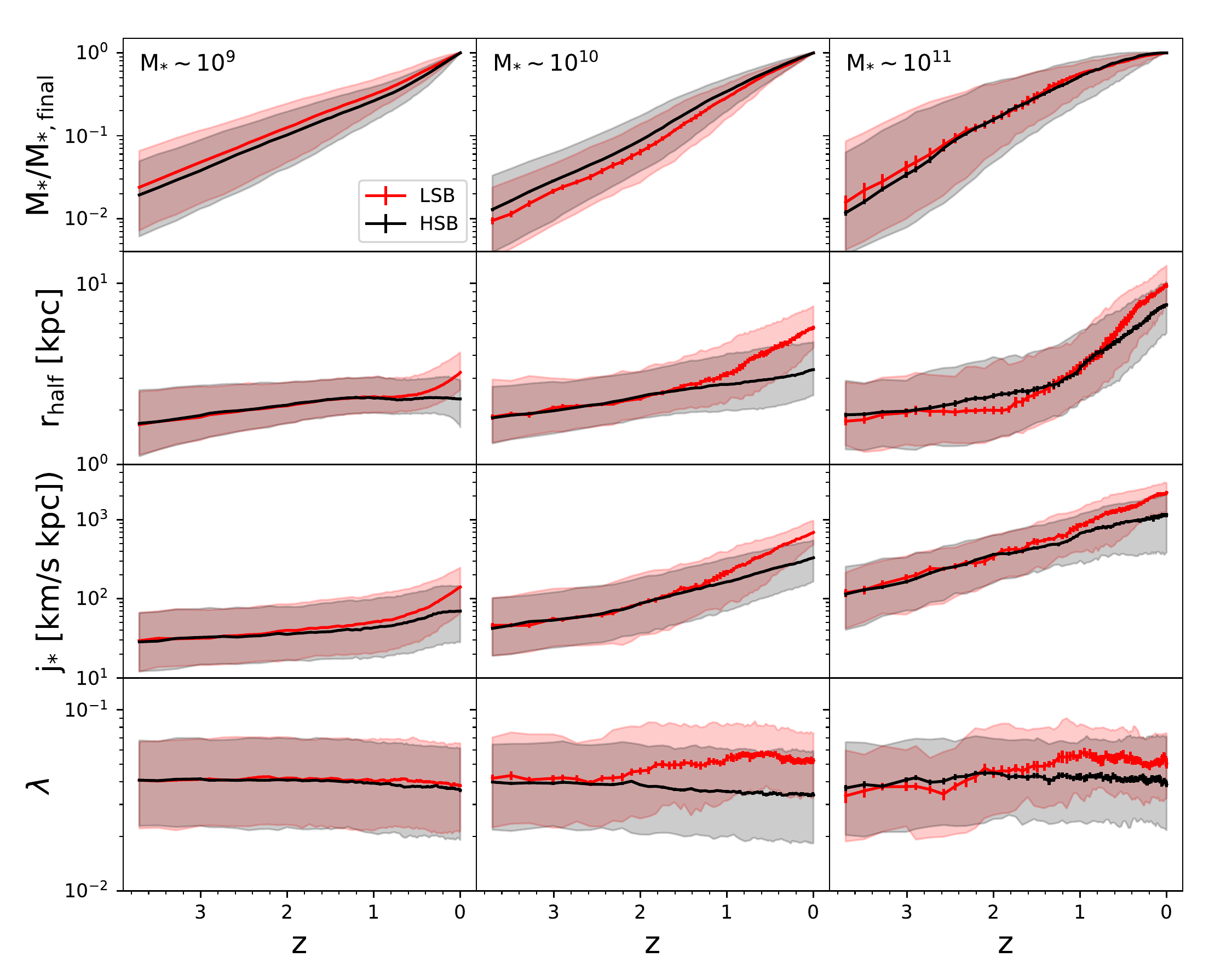} 
\caption{Median evolutionary tracks for the fraction of the final stellar mass (top row), galaxy size (second row), stellar specific angular momentum (third row) and halo spin (bottom row) for IllustrisTNG galaxies. The columns from left to right correspond to different stellar mass bins at $z=0$, corresponding to (0.7-1.4)$\times 10^{9,10,11}$ M$_{\odot}$. As before, the shaded regions indicate the 16th to 84th percentile ranges, while the red and black lines correspond to LSBGs and HSBGs, respectively. The progenitors of LSBGs become systematically larger (and acquire higher stellar specific angular momentum) than the progenitors of HSBGs at $z \sim 0.5$--$1.5$. The segregation in the halo spin parameters of LSBG and HSBG progenitors takes place at an earlier redshift, around $z \sim 2$.}
\label{fig:history_main}
\end{figure*}

\begin{figure*} 
\centering
\includegraphics[width=0.9\textwidth]{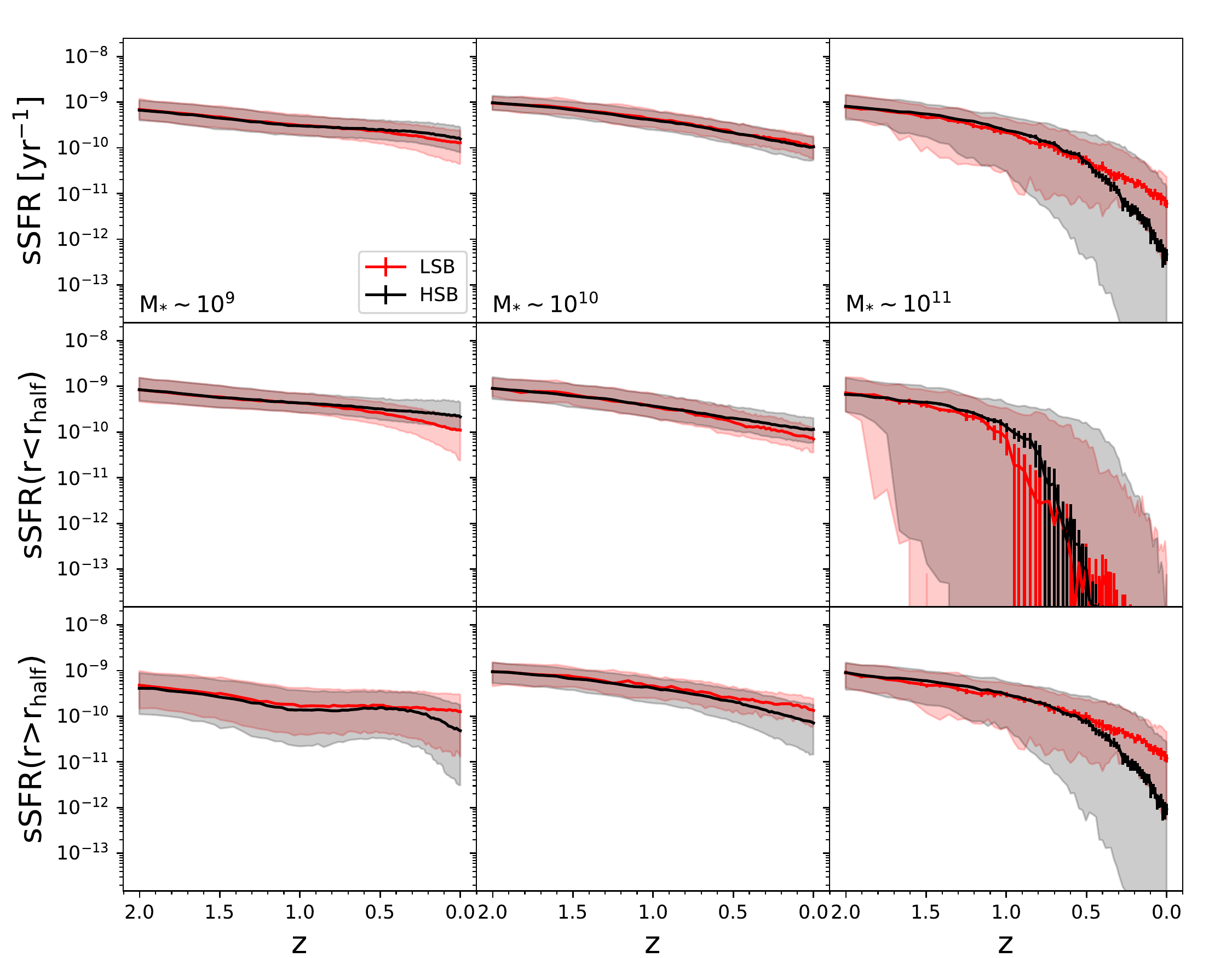}
\caption{Evolution of the total sSFR (top row), the sSFR within one effective radius (middle row), and the sSFR beyond one effective radius (bottom row), as indicated by the $y$-axis labels. The columns from left to right correspond to different stellar mass bins at $z=0$, corresponding to (0.7-1.4)$\times 10^{9,10,11}$ M$_{\odot}$. As found previously, we do no observe significant differences between the sSFRs for galaxies with $M_{\ast} \approx 10^{9}$--$10^{10}$ M$_\odot$, but we do observe that the sSFR tends to be less intense in the inner regions and more active in their outer parts of LSBGs, especially since $z\sim$ 1.5, where $r_{\rm half}$ becomes larger for LSBGs, supporting the idea that LSBGs form most of their stars in the outer parts of their discs.}
\label{fig:history_sfr_all}
\end{figure*}

\section{Redshift evolution}
\label{sec:lsb_evol}

	In this section we study the evolution of the galaxies in our sample using the \textsc{SubLink} merger trees \citep{RodGom15} for the LSBGs and HSBGs selected at $z=0$. We first track the redshift evolution of the main properties studied in Sections \ref{sec:global} and \ref{sec:angmom}, and then follow their resolved star formation histories (SFHs). Throughout this section we select galaxies at $z=0$ in three different stellar mass ranges, centred at $10^{9}$, $10^{10}$ and $10^{11}$ M$_\odot$.

\subsection{Masses, sizes and angular momenta}
\label{sec:mass_evol}
		Fig. \ref{fig:history_main} shows the median values of the instantaneous-to-final stellar mass ratio ($M_{*}$/$M_{\rm *,final}$, top row), the radius containing half of the total stellar mass (second row), the specific angular momentum of the stellar component (third row) and the halo spin parameter (bottom row) as a function of redshift for LSBGs (red lines) and HSBGs (black lines). The panels from left to right include galaxies selected according to their $z=0$ stellar mass in bins centred at $10^{9}$, $10^{10}$ and $10^{11}$ M$_\odot$.
		
		We can observe from the first row in Fig. \ref{fig:history_main} that there is no significant difference in the stellar mass formation history between LSBGs and HSBGs, implying that both kinds of galaxies may follow, in principle, similar stellar assembly processes. On the second row of Fig. \ref{fig:history_main}, we plot the median values of the redshift evolution of $r_{\rm half}$, which show a clear bifurcation at $z \sim 0.5$--$1.5$, where LSBGs begin to exhibit larger radii than HSBGs, resulting in the more extended radii of LSBGs clearly visible at $z =0$ (see Fig. \ref{fig:galaxy_prop}). When we examine the evolution of $j_{*}$ we find a similar divergence at $z \sim 1.5$ for all mass ranges, which in principle could be responsible of the increase of $r_{\rm half}$ for LSBGs.
	
	Finally, the fourth row of Fig. \ref{fig:history_main} shows the $\lambda$ evolution of the parent haloes of LSBGs and HSBGs, which follow a similar behaviour at early epochs and then exhibit a divergence of nearly a factor of 2 in their median values (as previously found by different authors, e.g., \citealt{PerezMontano19, Kulier20, Salinas21}), especially for galaxies with intermediate and high stellar masses. For low-mass galaxies, the differences in $\lambda$ become less noticeable, suggesting that other mechanisms might play a more important role \citep[such as supernovae-driven galactic outflows, e.g.][]{DiCintio19}. Interestingly, the bifurcation in $\lambda$ values occurs at a slightly earlier redshift, about $z \sim 2$, which suggests that the increase in the spin parameter may have a causal effect on other physical properties of the LSBGs, such as their stellar angular momentum and, consequently, their sizes. Note that the redshift of this bifurcation roughly corresponds to the `turnaround' epoch for these haloes \citep[e.g.][]{White84, Zavala16}, after which the halo spin parameter presents very weak evolution. 

\subsection{Star formation histories}
\label{sec:sfh}

To further understand the origin of the different present-day properties shown by LSBGs and HSBGs, in Fig. \ref{fig:history_sfr_all} we show their resolved star formation histories in different radial intervals. When we compared the sSFRs of LSBGs and HSBGs at $z=0$ in Section \ref{sec:gasSFR}, we did not find a clear difference between them, except for objects with stellar masses around $M_{\ast} \sim 10^{11} \, \Msun$. The panels in the top row from Fig. \ref{fig:history_sfr_all} confirm these results. However, given the more extended spatial distributions of LSBGs, we explore whether differences in the sSFR arise when considering the inner and outer regions of the galaxies in our sample. The middle row of Fig. \ref{fig:history_sfr_all} shows the redshift evolution of the sSFR measured \textit{within} $r_{\rm half}$ (i.e. the inner sSFR). Similarly, the bottom row shows the sSFR measured \textit{beyond} $r_{\rm half}$ (i.e. the outer sSFR).

We observe that, for galaxies with stellar masses between $\sim$10${^{9}}$ and $\sim$10${^{10}} \, \Msun$ (first two columns in Fig. \ref{fig:history_sfr_all}), the central sSFR (middle row) is less intense for LSBGs than for HSBGs at low redshift ($z < 1.5$). The opposite is observed when we look at the sSFR in the outer regions of the galaxies in our sample, as shown in the bottom row of the same figure, which indicates a more intense star formation in the outer regions of LSBGs when compared to HSBGs. We repeated the same analysis considering the regions within and beyond two times $r_{\rm half}$, finding a similar behaviour.

These findings are consistent with previous simulation studies \citep[e.g.][]{Zhu18, DiCintio19} where LSBGs are able to form through the accretion of high angular momentum material that promotes star formation at large radii. Our results also agree with those of \cite{Genel18}, who found that quenched galaxies at the massive end in TNG100 display a slower size growth, similar to our massive HSBG sample. Furthermore, \cite{Gupta21} found that massive, spatially extended galaxies at $z=2$ quench later than normal-sized galaxies, which is consistent with our finding that massive LSBGs retain a higher sSFR than HSBGs down to $z=0$.

\section{Discussion and Conclusions}
\label{sec:Conclusions}
	Using a sample of simulated galaxies from the TNG100 simulation of the IllustrisTNG project, we studied the main global properties of LSBGs at $z=0$ and identified their main differences with respect to HSBGs. We built our sample according to their face-on $r$-band effective surface brightness by considering the flux of those stellar particles within the projected radius containing half of the total $r-$band luminosity. Our galaxy sample included all TNG100 galaxies with $M_{*}$ $> 10^{9}$ M$_\odot$.
	
	Adopting a kinematic criterion to measure galaxy morphology, we found that LSBGs with stellar masses in the range 10$^{9}$--10$^{11}$ M$_\odot$ have a predominantly spiral morphology (Fig. \ref{fig:mu_vs_Mstar_kappa}). In our sample, the fraction of LSBGs is highest for low mass galaxies, with the fraction decreasing with increasing stellar mass, up to $M_{*}$ $ \sim 10^{10.5}$ M$_\odot$, where the fraction starts to increase. The high fraction of LSBGs at low stellar masses in our sample is in agreement with previous studies \citep{Dalcanton97,Galaz11}, while the upturn identified at intermediate masses coincides with the behaviour reported by \citet{Kulier20}, who attributed the formation of massive LSBGs to the accretion of an extended stellar component via mergers. 
	
  When analysing the distribution of key physical parameters of our sample at low redshift (Fig. \ref{fig:galaxy_prop}), we found that LSBGs have systematically fainter absolute magnitudes and lower SFRs than HSBGs. They are also less massive and more extended than HSBGs, as frequently reported by observational studies of LSBGs in the local Universe  \citep{Zhong08,Galaz11,PerezMontano19}.
    
    Once identified that the population of LSBGs is less massive than that of HSBGs, we explored the differences between the two populations at fixed stellar mass, in order to remove any dependence with $M_{*}$. By doing so, the difference in current star formation rate vanishes.
    
    The atomic, molecular and total neutral gas content is substantially higher for LSBGs, resulting in a higher baryonic-to-total mass ratio for galaxies in the low surface brightness regime. Although in the literature it is common to refer to LSBGs as DM-dominated \citep{Pickering97,McGaugh01,Swaters03,deNaray11,PerezMontano19}, the LSBGs of our sample present only a marginally smaller stellar-to-halo mass fraction when compared to HSBGs, which could be a result of our stellar and halo mass estimations, that correspond to the total stellar mass particles within the subhalo and the DM mass enclosed within $R_{200}$, physical properties not directly attainable from observations. On the other hand, estimations from observational studies are restricted to regions traceable by a baryonic component, which are typically much smaller than $R_{200}$.
    
    The fraction of stellar mass accreted via mergers is slightly higher for low-mass LSBGs than for HSBGs, where the ex-situ stellar fraction contribution is less than 40\%. Massive LSBGs present also larger ex-situ stellar fractions than HSBGs, where the contribution is up to 80\%. This allows us to conclude that the extended nature of high-mass LSBGs is moderately affected by accreted stellar mass. We also find that, for a restricted range of stellar masses, between $10^9$ and $10^{10}$ M$_\odot$, LSBGs experienced their last major merger more recently, indicating that these assembly modes might be relevant in the formation of LSBGs (Fig. \ref{fig:history_prop}). For instance, \citet{DiCintio19} proposed that coplanar mergers are able to decrease the surface brightness of galactic discs by producing merger remnants with higher angular momentum. Other theoretical studies have shown that interactions of massive galaxies with less massive, gas-rich companions are able to produce giant LSBGs similar to Malin-1 (\citealt{Zhu18}; Zhu et al., in prep.).
    
    Two physical parameters that play a key role in the formation of LSBGs are the galactic angular momentum and the halo spin parameter (Fig. \ref{fig:angmom}), with LSBGs presenting substantially higher median values of the specific stellar angular momentum ($j_{\ast}$) and halo spin parameter ($\lambda$) than HSBGs, in line with classical works that propose the formation of LSBGs embedded in dark matter halos with high values of $\lambda$ \citep{Dalcanton97, Jimenez98, Boissier03, Salinas21}. Some recent hydrodynamic simulations find similar correlations with halo spin \citep[e.g.][]{Kulier20}, while others find weaker correlations \citep{Jiang2019} or even opposite trends \citep{Martin19}. Not only do LSBGs present higher $\lambda$ values, but they also present a higher `retention fraction' ($j_{\ast} / j_{200}$) of angular momentum compared to HSBGs. Hence, this high angular momentum configuration is responsible of the low surface brightness nature of these systems.
    
    Exploring the mass and accretion rate of the supermassive BHs at the centres of our galaxies, we found that the ones found in LSBGs are less massive and present lower accretion rates than the BHs hosted by HSBGs, a result that can be attributed, as proposed by \citet{Cervantes11}, to the high angular momentum and halo spin parameter exhibited by LSBGs, which would prevent the fuelling of material to the central regions. The lower accretion rates found in LSBGs compared to HSBGs at the low-mass end reflect the lower density environments at the centres of LSBGs. A similar result was recently found by \citet{RodGom22} using the TNG100 simulation, which they also attributed to less efficient gas inflow toward the galactic centre in systems with higher angular momentum. Another critical factor that could contribute to the low accretion rate and less massive BHs in LSBGs is the lower occurrence of stellar bars in these kind of galaxies \citep{Honey16, Cerv17}, as bars promote angular momentum and mass redistribution between the stellar, gas and dark matter components of the galaxies. In particular, bars promote the transfer of gas to the centre of the galaxy, material that could be used as fuel for the central BHs.

Exploring the redshift evolution of the galaxies in our sample (Fig. \ref{fig:history_main}), we found that LSBGs and HSBGs follow the same stellar mass formation history, but their size growth shows a clear bifurcation at $z \sim 0.5-1.5$, with the half-mass radius of LSBGs increasing at a higher rate. A similar bifurcation is present for the case of the specific stellar angular momentum and the spin parameter, but for the case of the spin the divergence occurs at higher redshift ($z \sim 2$), which suggests that the increase in the spin parameter may have a causal effect on other physical properties of LSBGs, such as $r_{\rm half}$, $j_{*}$, and the star formation rate.

In a recent study, \citet{Wright21} proposed that an increase in halo spin can cause a redistribution of the star formation to the outskirts of galaxies, reducing the central star formation rate, a phenomenon that might explain, at least in part, what we find in our sample, where for galaxies at low redshift ($z < 1.5$) within $10^9-10^{10}$ M$_\odot$, the central sSFR is less intense for LSBGs than for HSBGs, and more intense in the outer regions (Fig. \ref{fig:history_sfr_all}). This extended star formation in turn increases the stellar radii of the LSBGs.

Altogether, our main results can be summarized as follows:

(i) LSBGs are less massive, more gas-rich, and more extended than HSBGs, and present later-type morphologies.

(ii) LSBGs have less massive BHs and lower BH accretion rates than HSBGs, likely due to their higher angular momentum, which prevents the material from falling into the central regions of the galaxy.

(iii) The DM haloes of LSBGs assemble half of their final mass in slightly more recent times, reaffirming, together with their younger stellar populations, the hypothesis that LSBGs are marginally younger objects.

(iv) At a fixed stellar mass, LSBGs consistently exhibit higher halo spin parameters and angular momentum `retention fractions' than HSBGs. The combination of these effects results in a higher stellar specific angular momentum.

(v) At fixed stellar mass, LSBGs are formed within marginally more massive DM halos with larger amounts of baryonic material than those of HSBGs, mainly in the form of non-star-forming gas.

(vi) The values of $\lambda$ and $j_{\ast}$ diverge at $z\sim 2-1.5$ and, as a consequence, the star formation activity seems to be affected \citep[see also][]{Gupta21}. Around $z \sim 1 - 0.5$, LSBGs and HSBGs have similar star formation histories, but then both populations diverge, especially when we compare the star formation activities in their inner and outer regions. LSBGs have had most of their star formation activity in their outer regions.

In summary, our results indicate that LSBGs tend to be systems with high amounts of angular momentum and cold gas, probably as a result of the higher spin parameter of their host haloes, as predicted in the classical framework of galaxy formation for these kind of galaxies. While the total sSFRs of LSBGs and HSBGs do not differ significantly, the spatial distribution of the sSFR is more extended in the case of LSBGs, being higher in the outer regions when compared to HSBGs. Although we find that mergers also favour the formation of LSBGs, their effect becomes more important at the massive end, in the regime of so-called \textit{giant} LSBGs (e.g. \citealt{Zhu18}, Zhu et al., in prep.). On the other hand, environmental effects are believed to be more important in the formation of the so-called \textit{Ultra Diffuse Galaxies} (UDGs) (e.g. \citealt{Tremmel2020, Benavides2021}), at even fainter surface brightnesses. In upcoming work, it will be interesting to explore the combined effect of mergers and environment on our sample of LSBGs. Finally, our results motivate future studies of LSBGs using deeper galaxy surveys, such as Euclid and LSST, as well as more direct comparisons between observations and simulations using synthetic observations.

\section*{Acknowledgements}
We thank Gaspar Galaz and Miguel Aragón-Calvo for useful comments and discussions during the early phases of this project, and Yetli Rosas-Guevara for comments on the final results. We are also grateful to the referee, Yannick Bahé, for an insightful report that helped to improve the quality of the paper.
Bernardo Cervantes Sodi and Luis Enrique P{\'e}rez-Monta{\~n}o acknowledge financial support through Programa de Apoyo a Proyectos de Investigación e Innovación Tecnológica (PAPIIT) project IA103520 from Dirección General de Asuntos del Personal Académico de la UNAM(DGAPA-UNAM).
The IllustrisTNG flagship simulations were run on the HazelHen Cray XC40 supercomputer at the High Performance Computing Center Stuttgart (HLRS) as part of project GCS-ILLU of the Gauss Centre for Supercomputing (GCS). Ancillary and test runs of the project were also run on the compute cluster operated by HITS, on the Stampede supercomputer at TACC/XSEDE (allocation AST140063), at the Hydra and Draco supercomputers at the Max Planck Computing and Data Facility, and on the MIT/Harvard computing facilities supported by FAS and MIT MKI.

\section*{Data availability}

The data from the IllustrisTNG simulations used in this work are publicly available at the website \href{https://www.tng-project.org}{https://www.tng-project.org} \citep{Nelson19a}.




\bsp	
\label{lastpage}
\end{document}